\pdfoutput=1
\documentclass[reprint,superscriptaddress,amsmath,amssymb,aip,apl]{revtex4-2}

\usepackage{graphicx}% Include figure files
\usepackage{dcolumn}% Align table columns on decimal point
\usepackage{bm}% bold math
\usepackage{soul,xcolor}
\usepackage[colorlinks=true,linkcolor=blue,urlcolor=blue,citecolor=blue]{hyperref}%
\usepackage{siunitx}

\begin{document}
\setstcolor{red}

\preprint{AIP/123-QED}

\title{Cavity-Dumping a Single Infrared Pulse from a Free-Electron Laser for Two-Color Pump-Probe Experiments}

\author{T.~Janssen}
\altaffiliation[]{Corresponding author; email: thom.janssen@ru.nl}

\affiliation{FELIX Laboratory, Radboud University, Toernooiveld 7, 6525 ED Nijmegen, The Netherlands\looseness=-1}%

\affiliation{Radboud University, Institute for Molecules and Materials, Heyendaalseweg 135, 6525 AJ Nijmegen, The Netherlands\looseness=-1}%

\author{C.~S.~Davies}
\affiliation{FELIX Laboratory, Radboud University, Toernooiveld 7, 6525 ED Nijmegen, The Netherlands\looseness=-1}%
\affiliation{Radboud University, Institute for Molecules and Materials, Heyendaalseweg 135, 6525 AJ Nijmegen, The Netherlands\looseness=-1}%

\author{M.~Gidding}
\affiliation{FELIX Laboratory, Radboud University, Toernooiveld 7, 6525 ED Nijmegen, The Netherlands\looseness=-1}%
\affiliation{Radboud University, Institute for Molecules and Materials, Heyendaalseweg 135, 6525 AJ Nijmegen, The Netherlands\looseness=-1}%

\author{V.~Chernyy}
\affiliation{FELIX Laboratory, Radboud University, Toernooiveld 7, 6525 ED Nijmegen, The Netherlands\looseness=-1}%
\affiliation{Radboud University, Institute for Molecules and Materials, Heyendaalseweg 135, 6525 AJ Nijmegen, The Netherlands\looseness=-1}%

\author{J.~M.~Bakker}
\affiliation{FELIX Laboratory, Radboud University, Toernooiveld 7, 6525 ED Nijmegen, The Netherlands\looseness=-1}%
\affiliation{Radboud University, Institute for Molecules and Materials, Heyendaalseweg 135, 6525 AJ Nijmegen, The Netherlands\looseness=-1}%

\author{A.~Kirilyuk}
\affiliation{FELIX Laboratory, Radboud University, Toernooiveld 7, 6525 ED Nijmegen, The Netherlands\looseness=-1}%
\affiliation{Radboud University, Institute for Molecules and Materials, Heyendaalseweg 135, 6525 AJ Nijmegen, The Netherlands\looseness=-1}%

\date{\today}

\begin{abstract}
Electromagnetic radiation in the mid- to far-infrared spectral range represents an indispensable tool for the study of numerous types of collective excitations in solids and molecules. Short and intense pulses in this THz spectral range are, however, difficult to obtain. While wide wavelength-tunability is easily provided by free-electron lasers, the energies of individual pulses are relatively moderate, on the order of microjoules. Here we demonstrate a setup that uses cavity-dumping of a free-electron laser to provide single, picosecond-long pulses in the mid- to far-infrared frequency range. The duration of the Fourier-limited pulses can be varied by cavity detuning, and their energy was shown to exceed \SI{100}{\micro \joule}. Using the aforementioned infrared pulse as a pump, we have realized a two-color pump-probe setup facilitating single-shot time-resolved imaging of magnetization dynamics. We demonstrate the capabilities of the setup first on thermally-induced demagnetization and magnetic switching of a GdFeCo thin film and second by showing a single-shot time-resolved detection of resonant phononic switching of the magnetization in a magnetic garnet. 
\end{abstract}

\maketitle

\section{\label{sec:level1} Introduction}

Fundamental studies involving light in the mid- to far-infrared (IR) and terahertz (THz) spectral ranges have been severely hampered by the scarcity of light sources delivering sufficiently short and strong pulses~\cite{THz-review}. Nonetheless, the terahertz gap houses a variety of interesting condensed matter physics phenomena, since the majority of materials feature characteristic excitations in this range. These collective modes or quasiparticles (phonons, magnons, phasons, spin fluctuations, cyclotron resonances etc.) belong to the realm of so-called low energy physics, an extremely important and intriguing energy range that determines all the thermodynamic and macroscopic properties of solids, such as ferroelectric, magnetic and crystallographic order, values of the critical temperature in superconductors and so on. Sufficiently strong excitation of these modes at the appropriate energy can induce transient metal-insulator transitions in manganite oxides driven purely by non-linear phononics~\cite{Rini}, extend the range of operational temperatures for superconductivity~\cite{Mitrano} and even lead to a permanent change in the order parameter~\cite{stupa1}. Such striking material changes, however, requires reaching into the nonlinear range, which in turn assumes an excitation of very high amplitude. In brief, short and intense pulses of light matching the characteristic frequencies of collective modes are required to drive the corresponding resonances. 

There are a number of ways to generate IR and THz light ranging from the excitation of vibrational transitions in CO$_{2}$~\cite{co2laser} and electronic transitions in a quantum cascade structure~\cite{QCL1,QCL2} to the non-linear mixing of optical pulses delivered by optical parametric oscillators~\cite{OPA1,OPA2}. The typical downside of these techniques, however, is the lack of large wavelength-tunability combined with a high output power.
In contrast, the development of the Free-Electron Laser (FEL) allows one to create short optical pulses across wavelength/intensity ranges that are otherwise difficult to reach. 
To generate such pulses, FELs accelerate electrons to relativistic speeds and force them through an alternating magnetic field~\cite{FEL1,FEL2}. The frequency of the emitted light is directly related to the period of the alternating magnetic field or so-called undulator period. 

The typical temporal structure of FEL radiation consists of so-called ``macropulses'', typically a few microseconds-long, filled with picosecond-long ``micropulses'' coming at a repetition rate ranging from tens of MHz to several GHz. This high repetition rate of the micropulses is often, however, unsatisfactory for studying optically-induced processes in solids. Stroboscopic pump-probe techniques, for example, rely on the system relaxing back to its initial equilibrium faster than the time-interval between consecutive pulses, and this condition is not met when significant thermalization needs to take place or if the pump induces a permanent change in the system. In addition, the energy of a single micropulse is often insufficient to reach the desired level of excitation, which is also somewhat related to the difficulty of tightly focusing such long-wavelength radiation. In such cases, it is compulsory to reduce the repetition rate of the IR laser towards the single-shot level, while simultaneously boosting its energy.

Here we present the development of an experimental setup that allows one to obtain single, picosecond-long pulses of radiation from the mid-IR to THz spectral range, with energy far exceeding \SI{100}{\micro \joule}. This is achieved by dumping single micropulses directly from within the laser cavity. For such cavity-dumping we use a thin silicon slab introduced into the laser cavity at the Brewster angle as a plasma switch~\cite{Sbarbara}. An application of a nanosecond-long pulse from an externally-synchronized frequency-doubled Nd:YAG laser transforms the otherwise transparent slab into a transient mirror, thus reflecting a single pulse. Using the cavity-dumped IR pulse as a pump, we have realized a time-resolved single-shot imaging setup with nanosecond resolution. The capabilities of the setup are demonstrated on the example of all-optical single-shot switching of magnetization in ferrimagnetic alloys and magnetic dielectrics. 

\section{\label{sec:level2} Experimental implementation}

\subsection{Suite of free electron lasers at FELIX}

The four FELs housed at the FELIX facility (name derived from ``Free Electron Laser for Infrared eXperiments'') deliver laser radiation in different spectral regimes within the IR spectrum, with wavelengths ranging in total from \SI{2.7}{\micro\meter} to \SI{1.5}{\milli\meter}. All the FELs generate 6-\SI{10}{\micro \second}-long bunches of optical pulses (``macropulses'') at a user-selected repetition rate of 1, 2, 5 or \SI{10}{\hertz}. The macropulses themselves consist of micropulses with a repetition rate ranging between \SI{25}{\mega\hertz} and \SI{3}{\giga\hertz} depending on which FEL and operation mode is used. As a result, each macropulse comprises at least 200 micropulses, each with an energy of about 10-\SI{30}{\micro \joule}, depending on the wavelength range. 

Unlike in the fast-growing population of X-ray FELs, the long-wavelength versions are constructed with a real optical cavity. Therefore, the usable intensity outside the cavity is many times smaller than that inside it, exactly as is the case with usual visible-range lasers. Inside the cavity, the micropulses typically have an energy on the order of millijoules, but with the traditional method of outcoupling light via a small hole in the downstream cavity mirror, only a fraction ($2\% - 5\%$) of the energy leaves the cavity~\cite{lex2}. 
Moreover, the energy that is delivered to the user-stations is lower still, because first a part (1\%) of the beam is sampled at the diagnostic station and second due to some losses (2\%) in the transport system~\cite{Lex3}. 

A technique of ``pulse-slicing'' has been previously developed at FELIX that allows for a single micropulse to be separated from a full macropulse~\cite{Lex1}. For this, a silicon plate is placed in the path of the outgoing IR beam under the Brewster angle. When the Si is excited with an intense laser with photon energy above the semiconductor's bandgap, a dense electronic plasma is generated for about \SI{10}{\nano \second}. Consequently, the otherwise-transparent semiconductor is transformed into a transient mirror and a single micropulse is reflected (``sliced'') from the macropulse~\cite{Lex1}. At FELIX, pulse-slicing is conventionally realized using a frequency-doubled Nd:YAG laser delivering \SI{5}{\nano\second}-long pulses with a photon energy of \SI{2.3}{\electronvolt} and a pulse energy of $\approx$\SI{100}{\milli\joule} in combination with a silicon plate (bandgap \SI{1.1}{\electronvolt}). At optimum conditions, single pulses with energy reaching up to \SI{12}{\micro \joule} can be obtained. 

To be able to exploit the full intensity inside the cavity, one of the lasers (named ``FELICE'' -- Free Electron Laser for Intra-Cavity Experiments) was designed with a nine-meter-long folded cavity that primarily facilitates intra-cavity experiments with gas-phase molecules and clusters~\cite{FLC1}. The FELICE laser has its own dedicated undulator and a four-mirror cavity that runs up through the ceiling of the vault towards a dedicated experimental area where one can directly access the cavity for experiments. In the operation mode that is best suited for cavity-dumping, FELICE generates $\approx$ \SI{8}{\micro \second} long macropulses at \SI{10}{\hertz} consisting of micropulses with a repetition rate of \SI{16.667}{\mega \hertz}.

Therefore in this work we introduce the pulse-slicing setup that is implemented directly inside the cavity of FELICE to select single infrared/THz pulses. We note that similar cavity-dumping schemes have been previously developed at the millimeter-wave FEL at the University of California Santa Barbara~\cite{Sbarbara} and the FELBE FEL~\cite{GeAu}.

\subsection{Operation of FEL with Si plate inside the cavity}

%%%%%%%%%%%%%%%%%%% fig 1 %%%%%%%%%%%%%%%%%%%%%%%%
\begin{figure}[b]
    \centering
    \includegraphics[width = 0.5\textwidth]{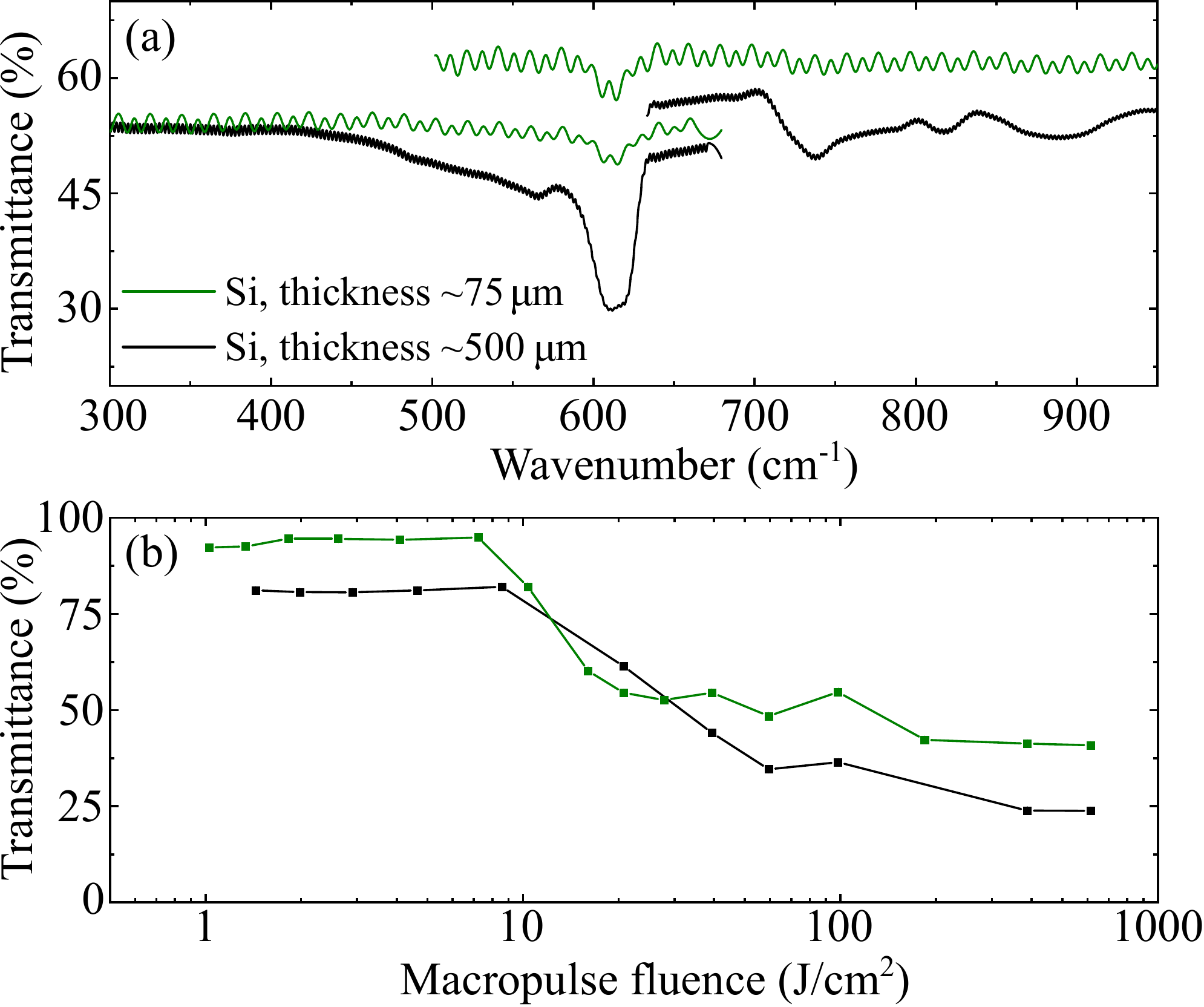}
    \caption{(a) The FTIR transmission spectrum recorded for Si wafers of two different thicknesses, using a Bruker Vertex-v50 spectrometer. (b) The transmittance of the \SI{75}{\micro\meter}- and \SI{500}{\micro\meter}-thick Si slab as a function of the IR fluence for the wavenumber \SI{742}{\centi\meter^{-1}}.}
    \label{fig:FTIR}
\end{figure}

As the first step, a variety of semiconductor materials were evaluated with consideration of several properties that are important when used as a transient mirror in the cavity~\cite{Valeriy}. First, the transmission in the transparent ``off'' state should be as high as possible to minimize the cavity losses during the laser's buildup. Second, the material must be able to withstand the high intra-cavity fluence. Third and finally, the electronic plasma that creates the reflective ``on'' state in the material must decay within a time scale shorter than the \SI{60}{\nano \second} time-interval between two consecutive micropulses. As a result of this evaluation~\cite{Valeriy}, silicon proved to be the best material for our purpose. Compared to ZnSe, GaAs and CdSe, Si has the highest photo-induced reflectivity of 55\% in the required IR range and is able to withstand the required level of fluence inside the cavity.

%%%%%%%%%%%%%%%%%%%%% fig 2 %%%%%%%%%%%%%%%%%%%%%%%%%
\begin{figure}[b]
    \centering
    \includegraphics[width = 8 cm]{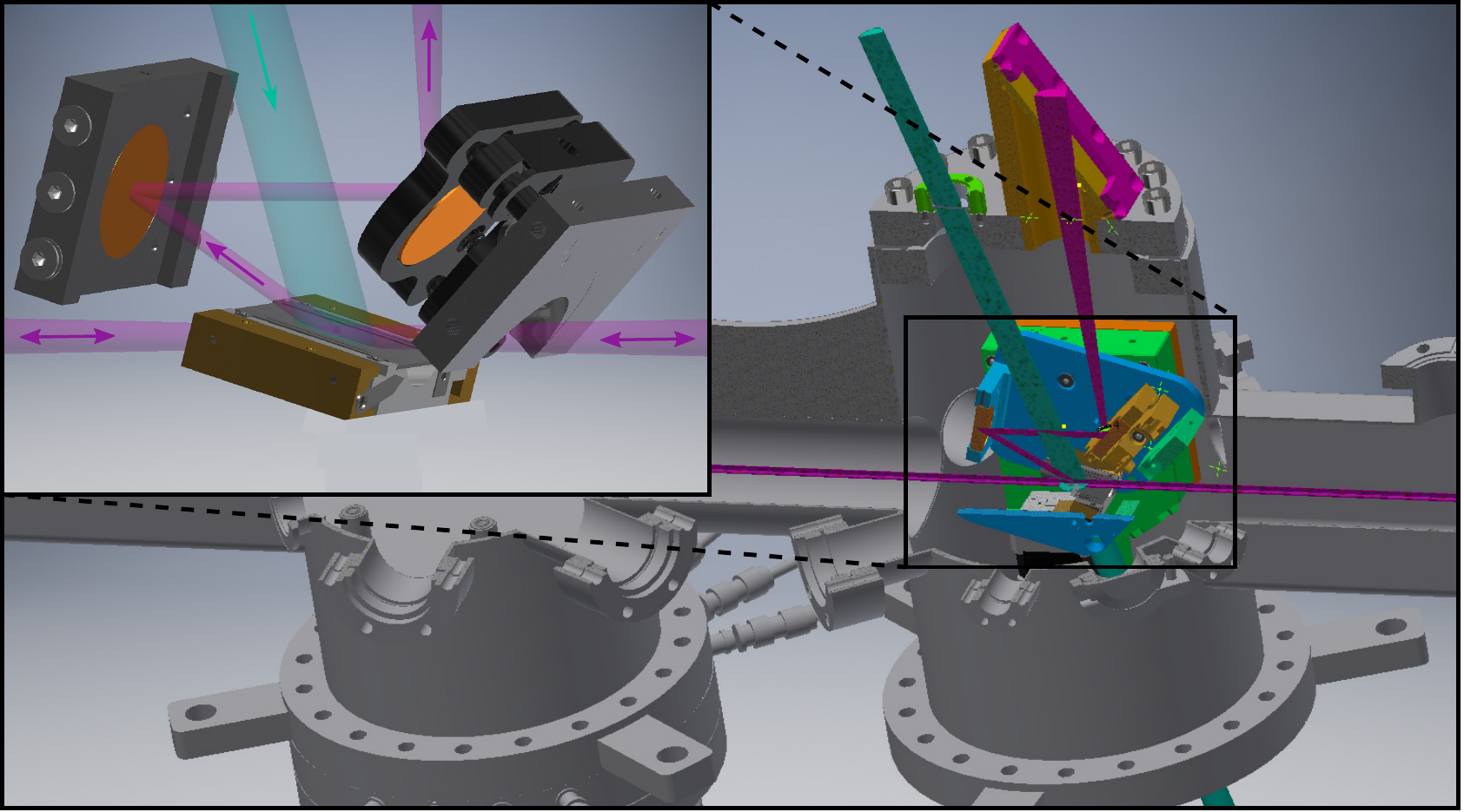}
    \caption{A schematic representation of the mirror assembly used for cavity-dumping. The optical paths of the Nd:YAG and IR pulses (green and purple respectively) intersect on a silicon slab placed in a removable tray. Inset: mirror assembly placed in the FELICE cavity.}
    \label{fig:setup}
\end{figure}

To achieve cavity-dumping, a mirror assembly containing a \SI{70}{\micro \meter}-thick double-side-polished crystalline silicon slab with crystal orientation (001) was designed and placed inside the FELICE cavity. Figure~\ref{fig:setup} shows the complete assembly including the Si slab positioned approximately at the waist of the IR beam (shown in purple). The entire assembly is mounted on a motorized stage and can be rotated in to and out of the path of the beam. In principle, the slab is placed at the Brewster angle with respect to the IR laser, allowing the buildup of laser energy inside the cavity to continue uninterrupted. At a suitable time, the slab is then illuminated with a pulse delivered by a frequency-doubled Spectra Physics Quanta-Ray INDI Nd:YAG laser (dark green in Fig.~\ref{fig:setup}) that is synchronized to FELICE's micropulse using a Quantum Composers 9520 Digital Pulse Delay Generator. The extracted IR pulse leaves the vacuum chamber through a ZnSe window also placed under Brewster angle. The window material limited the wavelength in our current experiments to about \SI{18}{\micro \meter}; replacing the window with e.g. KRS-5 will extend the usable range into much longer wavelengths.

\begin{figure}[t]
    \centering
    \includegraphics[width = 0.45\textwidth]{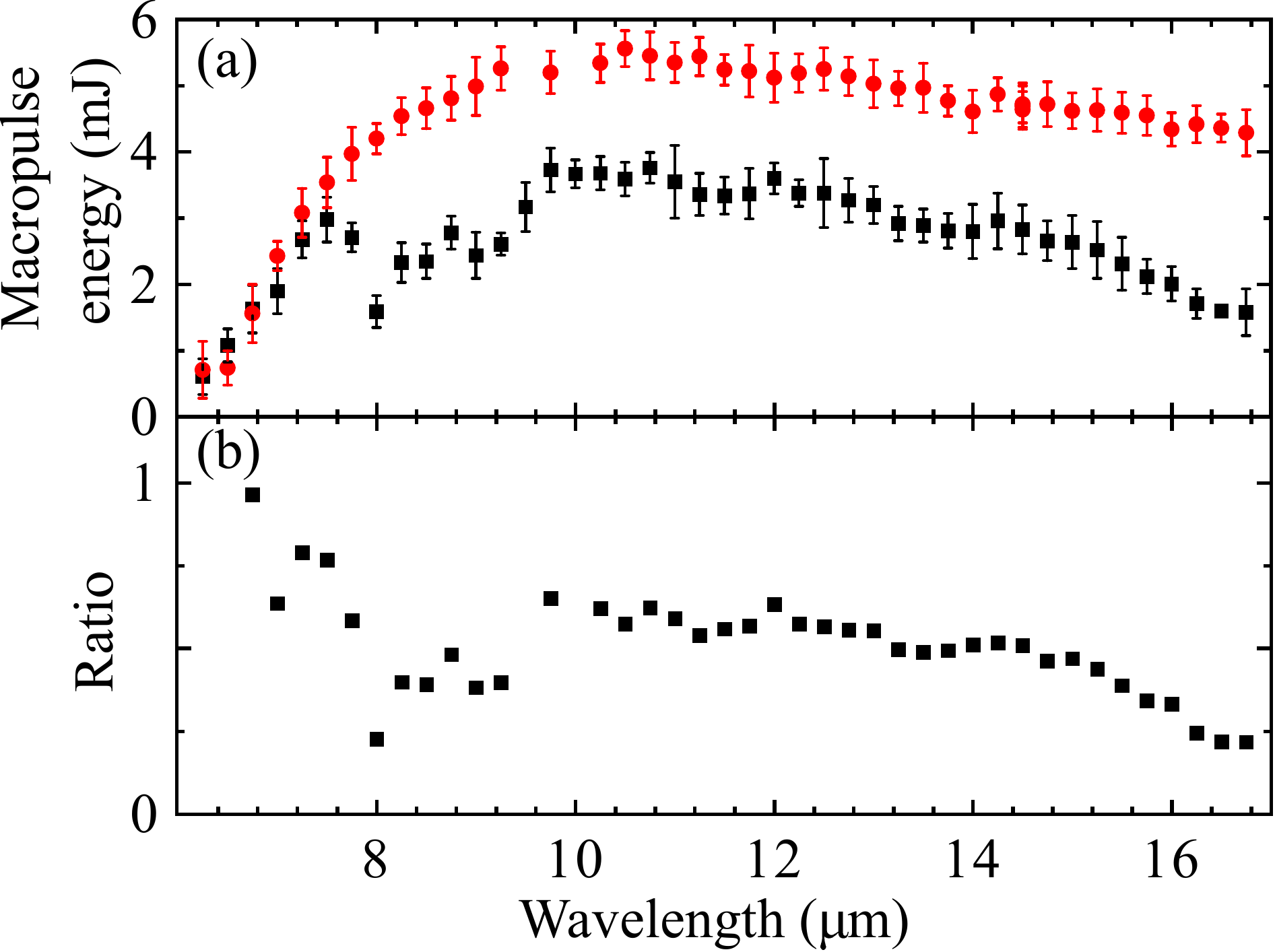}
    \caption{(a) Average outcoupled energy over 25 macropulses measured at the diagnostic station without (red points) and with (black points) the silicon slab inserted inside the cavity. (b) Ratio of FELICE outcoupled macropulse energy with the Si slab inside and outside the cavity.}
    \label{fig:Ratio}
\end{figure}

Under normal operating conditions, FELICE's losses amount to roughly 6\% per round-trip, 1\% of which is outcoupled for diagnostics.
The remainder of the losses is attributed to the four cavity mirrors. After placing the silicon slab inside the cavity the total power of FELICE was reduced by roughly 50\%, as shown in Fig.~\ref{fig:Ratio}. This significant reduction is explained by the very high light intensity inside the cavity and resulting multi-photon absorption processes in Si, as already revealed in Fig.~\ref{fig:FTIR}(b). Nevertheless, the laser remained in stable operation mode. Another aspect to consider when inserting the silicon slab in to the cavity relates to the fact that the effective optical path length of the cavity is increased, due to the much higher refractive index of silicon compared to the otherwise uninterrupted vacuum. In order to keep the cavity resonant for the given wavelength, FELICE is started with the slab retracted. As soon as it reaches stable operating conditions, the slab is inserted and the cavity is shortened by \SI{220}{\micro \meter}. After this, the lasing typically recovers automatically.

%%%%%%%%%%%%%%%%%%%%%% fig 3 %%%%%%%%%%%%%%%%%%%%%%%%%

\subsection{Cavity-dumping in operation}

\begin{figure}[t]
    \centering
    \includegraphics[width = 8cm]{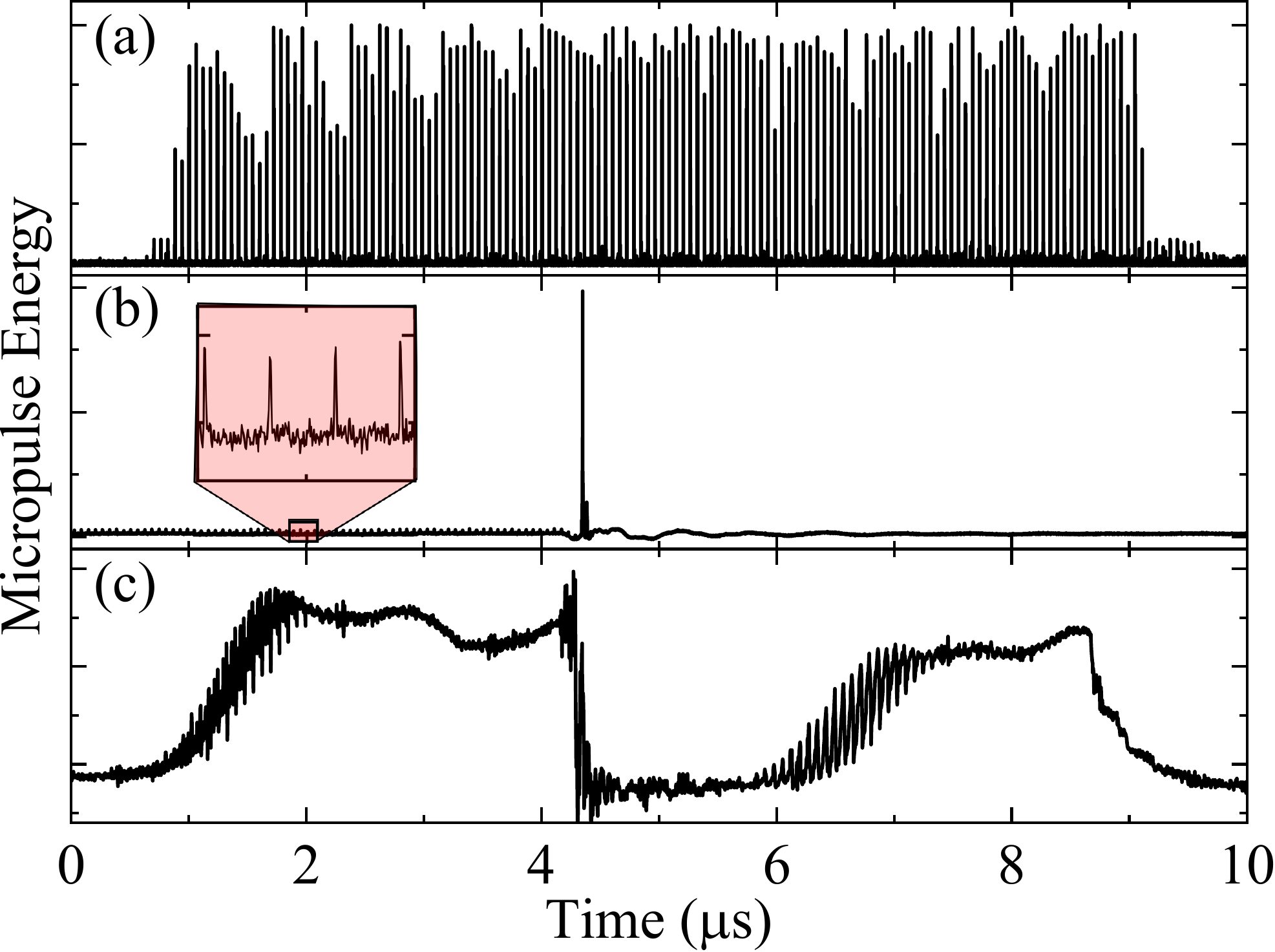}
    \caption{(a) A trace of the entire residual macropulse inside the laser cavity. (b) Single cavity-dumped pulse with \SI{15}{\decibel} attenuation in place. The inset shows the residual reflections magnified by a factor of 25. The ratio of energies between that of the dumped pulse and one of the residual reflections is 1:340. (c) Temporal trace of the macropulse outcoupled at the diagnostic station measured during the process of cavity-dumping. In this measurement, the Nd:YAG pulse reaches the slab at around \SI{4}{\micro \second} and destroys lasing in the cavity for about \SI{2}{\micro \second}.}
    \label{fig:Dump}
\end{figure}

To select a single micropulse from the macropulse, it is necessary to sample the timing of the macropulse inside the cavity. We achieve this by exploiting the fact that the Si wafer is imperfectly oriented at Brewster angle due to the focusing of the IR beam inside the laser cavity. Consequently, small (so-called ``residual'') reflections of the macropulse from inside the cavity can be detected using a high-bandwidth HgCdTe detector, as shown in Fig.~\ref{fig:Dump}(a). While Fig.~\ref{fig:Dump}(a) shows a large variety of micropulse intensity across the macropulse due to the weakness of the detected residuals, the micropulse intensity is actually fairly constant across the entire macropulse, as evidenced by the fact that it is lasing.

Figure~\ref{fig:Dump}(b) shows a trace of a single pulse extracted via cavity-dumping, but with an additional \SI{15}{dB} of attenuation in place compared to the measurement shown in Fig.~\ref{fig:Dump}(a). Because of the attenuation, residual reflections are not clearly visible, with the dumped micropulse being about 340 times stronger than one of the residual reflections.

%%%%%%%%%%%%%%%%%%%% fig 4 %%%%%%%%%%%%%%%%%%%%%%%%%%

Note that during operation in low repetition-rate mode, a single optical pulse is circulating in the laser cavity, with the final repetition rate (\SI{16.667}{\mega \hertz}) given by the cavity length (\SI{9}{\meter}). Therefore, adding the transient mirror inside the cavity and coupling a pulse out of it naturally destroys the operation of the laser, as illustrated in Fig.~\ref{fig:Dump}(c). At time $t=\SI{4.3}{\micro\second}$, the lasing of FELICE instantaneously (i.e. within a single period of \SI{60}{\nano\second}) collapses, which is followed, after more than a microsecond pause, by a slow (about \SI{2}{\micro\second}-long) recovery of the operation. This microsecond timescale of recovery corresponds to the regular build-up time of the FEL radiation. Since the micropulse intensity is rather constant within the macropulse, the exact timing of the Nd:YAG pulse (i.e. which of the micropulses is selected) is not important, and results in a dumped pulse of approximately the same energy. The energy of the single pulse only noticeably decreases when cavity-dumping at the very beginning or end of the macropulse.

\begin{figure}
    \centering
    \includegraphics[width = 6 cm]{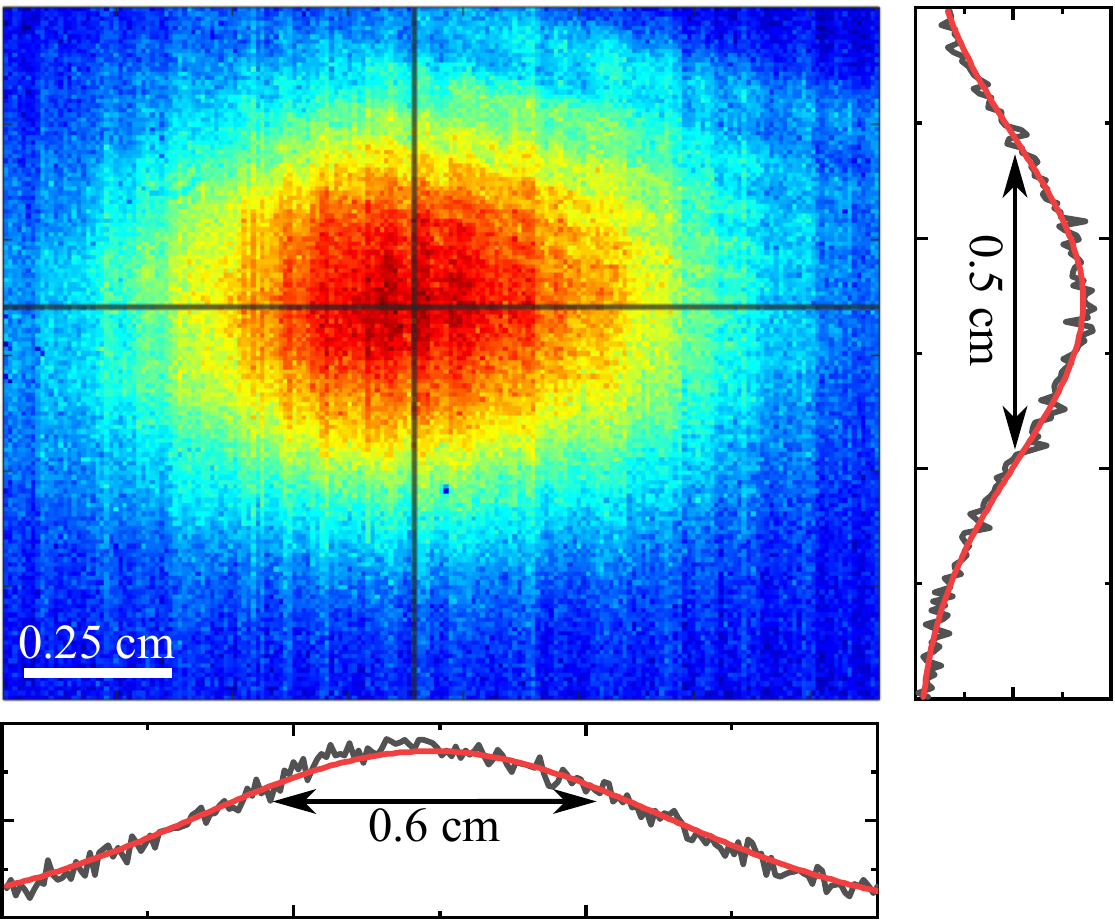}
    \caption{Beam profile of the dumped micropulse taken just after the out-coupling window. The intensity profile along the \textit{x}- and \textit{y}-axis is also plotted together with corresponding Gaussian fits.}
    \label{fig:2D}
\end{figure}

We assessed the spatial profile of the single cavity-dumped pulse using an infrared beam profiler (Spiricon Pyrocam IV). As expected from the position and orientation of the transient mirror relative to the focal point of the cavity and direction of propagation, the extracted pulse is significantly divergent, with an angle of $\Theta~=$~\SI{20}{\milli\radian} along the horizontal and vertical axes. The pulse is also slightly elliptical, probably due to the large tilt angle of the slab in the cavity. We therefore direct the micropulse onto a spherical mirror with a \SI{2}{\meter} radius of curvature, placed at the corresponding distance from the estimated beam waist inside the cavity, thus achieving a collimated beam with Gaussian distribution and a reduced divergence angle of $\Theta_{x}~=$~\SI{3.5}{\milli\radian} and $\Theta_{y}~=$~\SI{2.4}{\milli\radian}. Note that these values are close to the diffraction limit for the beam size of 5-\SI{7}{\milli\meter} (as can be seen on the profile shown in Fig.~\ref{fig:2D}) and wavelength of \SI{15}{\micro\meter}.  With the spherical mirror in place, the dumped pulse thus fits entirely on our 2-inch optics by the time it reaches the experimental area.

%%%%%%%%%%%%%%%%%%%%%% fig 5 %%%%%%%%%%%%%%%%%%%%%%%%%

\section{\label{sec:level3} Characterization of cavity-dumped pulses}

\begin{figure}
    \centering
    \includegraphics[width = 8.1cm]{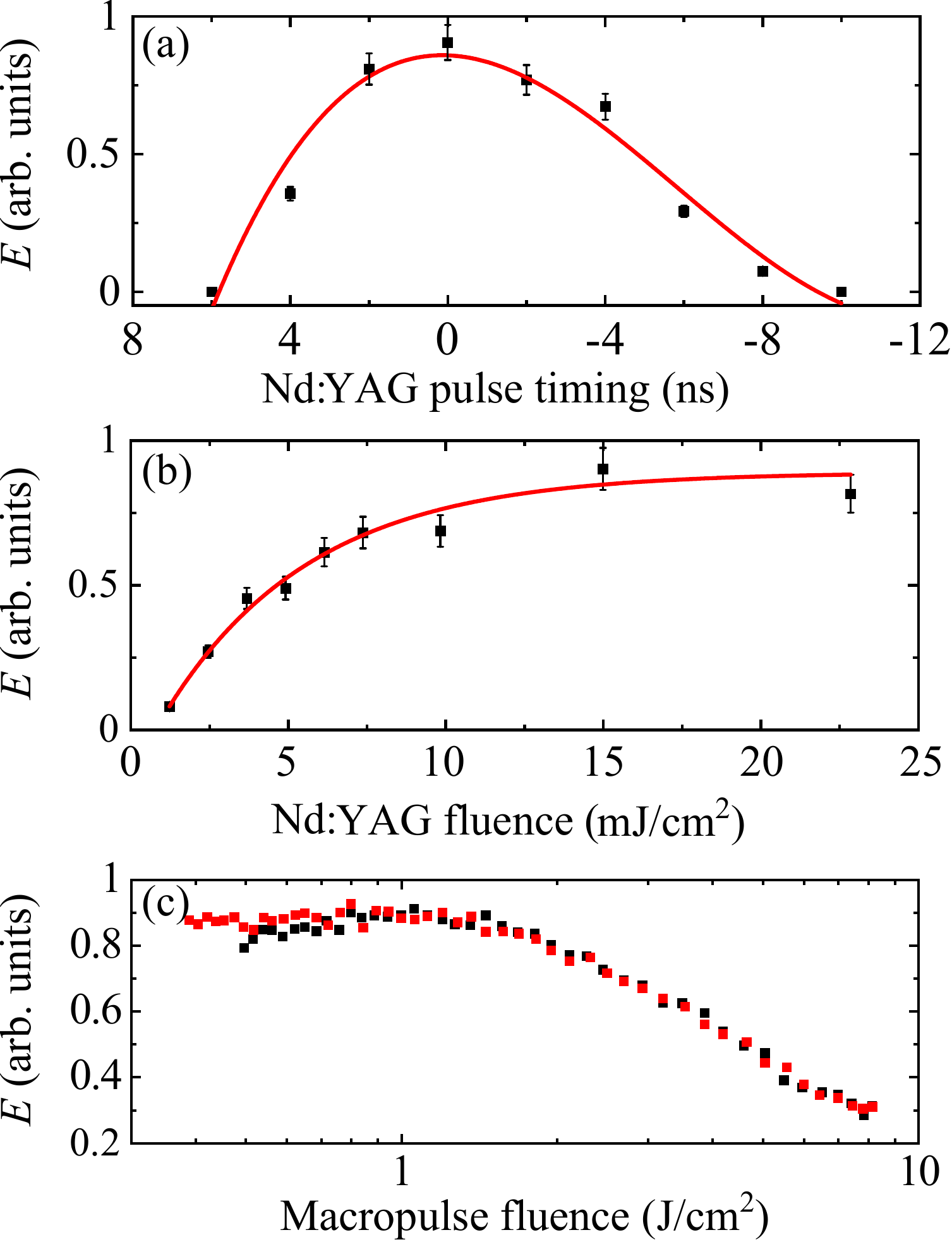}
    \caption{(a) Energy (\textit{E}) of the dumped micropulse as a function of the delay of the Nd:YAG laser pulse with respect to the cavity-dumped IR micropulse. (b) Energy of the dumped pulse for various fluences of the Nd:YAG pulse at a fixed time delay. (c) Energy of the dumped pulse for various fluences of the macropulse incident on the silicon slab, obtained by shifting the Si slab along the path of the focusing IR beam. In panels (a)-(b), the red line serves as a guide to the eye. The red and black points shown in panel (c) correspond to different measurements, showing the reproducibility of the energy-measurement.}
    \label{fig:YAG}
\end{figure}

The efficiency of the cavity-dumping process depends on several experimental parameters, which will be discussed in this section. The temporal overlap of the $\sim$~\SI{5}{\nano \second}-long Nd:YAG pulse and the individual micropulse on the silicon slab is the most important. Figure~\ref{fig:YAG}(a) shows a window of roughly \SI{14}{\nano \second} of time delay between the two pulses, where the micropulse is (partly) reflected off the illuminated silicon slab. ``Time-zero'' here is defined as the point where the strength of the reflected pulse is maximized. As the Nd:YAG pulse is scanned in time with respect to the micropulse, the reflectivity of the slab builds up over the course of roughly \SI{5}{\nano \second} and decays within \SI{10}{\nano \second}.

%%%%%%%%%%%%%%%%%%%% fig 6 %%%%%%%%%%%%%%%%%%%%%%%%%

The second important parameter is the energy of the Nd:YAG pulse. Figure~\ref{fig:YAG}(b) shows the strength of the dumped micropulse as a function of the Nd:YAG pulse energy. The reflectivity saturates when the Nd:YAG pulse energy reaches approximately \SI{60}{\milli \joule} per pulse. The size of the Nd:YAG laser spot was measured to be approximately \SI{3}{\centi \meter^2}, leading to an average optical fluence of around \SI{20}{\milli\joule\per\centi\meter^{2}} which is well under the estimated damage threshold of \SI{4.8}{\joule\per\centi\meter^2} of silicon measured in Ref.~[\onlinecite{dmgthres}]. A lens was added to investigate whether focusing of the Nd:YAG laser would increase the induced reflectivity, but the maximum energy of the dumped pulse was unaffected. From this, we confirm that the saturation of the plasma switch is already achieved using an unfocused beam.

Next we evaluated the influence of slab position with respect to the focal point of the cavity. For this, the whole setup was moved along the cavity, so that pulses from various positions along the converging FELICE beam inside the cavity were extracted. Shifting the position of the slab along the length of the cavity brings the slab closer to the focal point of the laser cavity, which is equivalent to increasing the macropulse fluence incident on the slab. In Fig.~\ref{fig:YAG}(c), we plot the energy of the cavity-dumped pulse as a function of the macropulse fluence incident on the slab. Close to the focal point of the IR laser, the slab is exposed to a macropulse fluence of $F~=~$\SI{8}{\joule\per\centi\meter^2} - here, the efficiency of the cavity-dumping is actually minimized because of the partial absorption resulting from the increase of multiphoton processes at these high intensities. As we move the slab away from the focal point, the macropulse fluence reduces but the cavity-dumped micropulse increases in energy. For the macropulse-fluence $F~=~$\SI{1}{\joule\per\centi\meter^2} (corresponding to the slab being \SI{160}{\milli \meter} away from the cavity's focal point), the cavity-dumped pulse maximizes in energy. Therefore we decided to fix the position the slab at a distance of \SI{150}{\milli \meter} away from the cavity's focal point for all further measurements and experiments.

The pulse duration of the micropulses is another important parameter as it may considerably affect the multi-photon processes in silicon, and thus reduce the performance of the setup. The pulse duration could be directly tuned by slightly changing the length of optical resonator, which results in a small change of phase slippage between the electron bunches and light pulses. While the duration can be changed by almost an order of magnitude between hundreds of femtoseconds and several picoseconds, the pulses stay Fourier-transform limited~\cite{Lex1}. Thus we verified that for all values of detuning used with our FELICE laser, the operation of cavity dumping setup is stable and pulses in a very broad duration/bandwidth range can be used.

\section{\label{sec:level4}Application of cavity-dumped pulses to time-resolved magnetic switching}

We further demonstrate the strength of the obtained cavity-dumped single pulses by studying the all-optical switching (AOS) of magnetization in two different systems. The first are metallic alloys of GdFeCo where thermally-induced switching was demonstrated previously~\cite{gd0} and subsequently studied in considerable detail~\cite{RPP}. Second, we test magnetic dielectrics where AOS was recently shown to be driven by the resonant excitation of optical phonons~\cite{stupa1}. In order to probe the switching in a time-resolved manner, we have developed a time-resolved single-shot imaging technique using $\sim$~\SI{5}{\nano \second}-long pulses derived from a second synchronized frequency-doubled Nd:YAG laser operating at \SI{20}{\hertz}.

\subsection{All-optical magnetic switching in GdFeCo}

The process of all-optical single-shot magnetic switching in GdFeCo alloys is driven purely by fast and efficient thermal demagnetization~\cite{gd1}, proceeding via different pathways of angular momentum flow depending on the pulse duration~\cite{gd2,gd3}. As equilibrated AOS in these samples has been previously intensively studied in the IR range~\cite{gd1}, it provides a good test for directly benchmarking the efficiency of the cavity-dumping and finding the temporal overlap of the IR pump and visible probe pulses at the sample position. For the excitation, single IR pulses were focused on the surface of the metallic film. The effect of the pump pulse on the sample magnetization is monitored at room temperature using a magneto-optical microscope sensitive to the out-of-plane component of magnetization. Specifically, the visible linearly-polarized probe pulses - of central wavelength 532~nm - illuminate the sample at an angle close to normal incidence, and the transmitted light is collected by an objective lens. An analyzer filters the rotation of the light's polarization due to the magneto-optical Faraday effect, and the image is recorded using a charge-coupled device.

\begin{figure}[t]
    \centering
    \includegraphics[width = 0.5\textwidth]{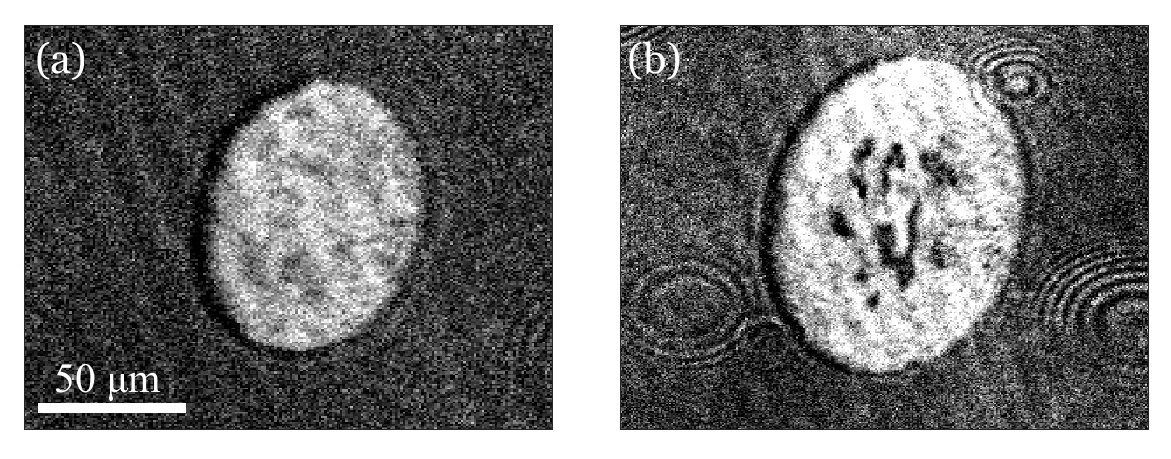}
    \caption{Two magneto-optical images showing the effect of a single infrared pulse of wavelength \SI{14.8}{\micro\meter} on Gd$_{27}$(FeCo)$_{73}$: (a) Magnetization reversal in the center of the affected area. (b) Demagnetized center and magnetization reversal at the edge of the affected area.}
    \label{fig:Gd27}
\end{figure}

As expected from the nominally Gaussian profile of the cavity-dumped IR pulse (Fig.~\ref{fig:2D}), the IR spot on the sample has a spatially-Gaussian distribution of energy. It is well-understood that single-shot all-optical switching of magnetization is subject to several thresholds in energy~\cite{Khorsand}. If the pulse energy is tuned to just above the threshold required for switching, a single spot of magnetization will reverse completely, as shown in Fig.~\ref{fig:Gd27}(a). If there is an excess of energy, however, the ferrimagnet becomes demagnetized due to the complete destruction and subsequent randomized recovery of magnetization. The Gaussian distribution of the IR pulse therefore frequently leads to the switching of magnetization at the outer perimeter of the irradiated region (where the local energy is sufficient for AOS) and demagnetization i.e. randomized domains at the center of the spot (where the local energy is too high), as shown in Fig.~\ref{fig:Gd27}(b). 

%%%%%%%%%%%%%%%%%%%%%% fig 7 %%%%%%%%%%%%%%%%%%%%%%%

Comparing single-pulse experiments from FELICE with those from FELIX~\cite{gd1} provides us with an estimate of the cavity-dumped pulse energy. During single-pulse experiments at the FELIX beamline, the GdFeCo sample shows a combination of switching and demagnetization when exposed to the full energy of the pulses obtained after slicing, similar to that shown in Fig.~\ref{fig:Gd27}(b). A clean magnetization reversal was typically achieved using 3 to \SI{5}{\decibel} of attenuation. In contrast, the use of cavity-dumped pulses allowed us to surpass the same energy threshold required for switching with typically \SI{18}{dB} (and sometimes even with \SI{20}{dB}) attenuation in the beam, indicating that the cavity dumped pulses are at least 30 times stronger than those obtained by pulse-slicing from FELIX. Compared with typically 3-\SI{5}{\micro \joule} energy in the FELIX-sliced pulses that reached the experimental setup, we thus roughly estimate the energy of cavity-dumped pulses to be on the order of 90-\SI{150}{\micro \joule}. To confirm these values, we used a power meter (Coherent Energymax J-10MB-LE) suited for the wavelength range of FELICE and which in principle is able to resolve single pulses. The maximum pulse energy measured for the cavity-dumped pulses was in excess of \SI{130}{\micro \joule} at a wavelength of \SI{14.8}{\micro\meter}. 

\begin{figure}[t]
    \centering
    \includegraphics[width=8cm]{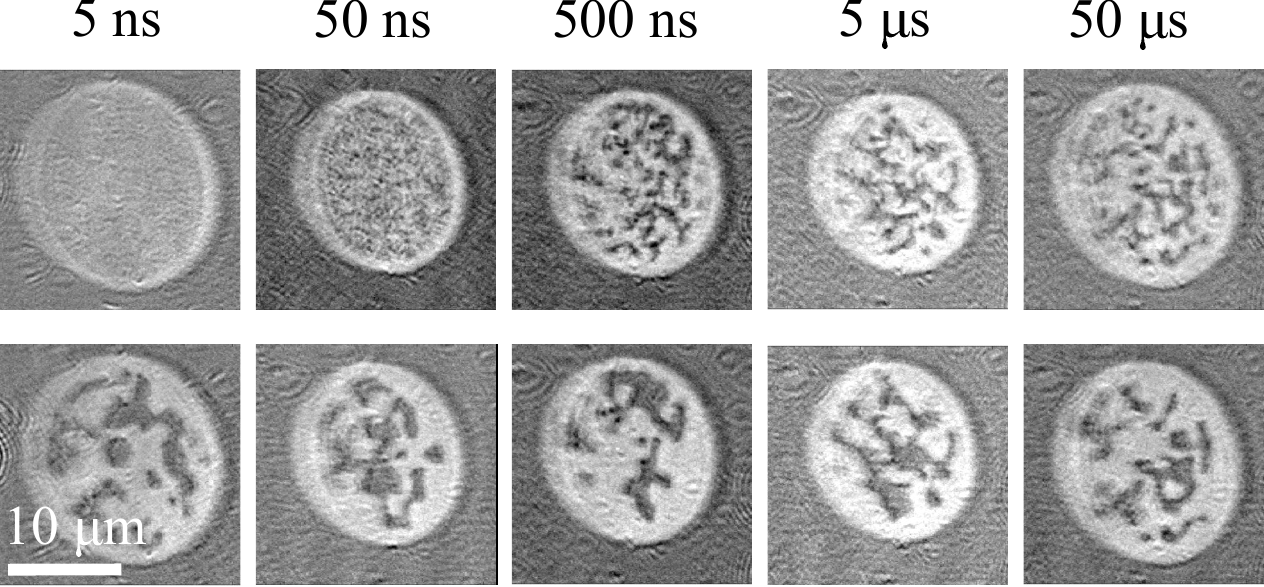}
    \caption{Magneto-optical images of the time-resolved process of demagnetization of Gd$_{27}$(FeCo)$_{73}$ pumped by a single IR micropulse of wavelength \SI{14.8}{\micro\meter}. The top row shows the time-resolved images showing the nucleation and growth of domains after laser excitation, and the bottom row shows the sample after full relaxation for the same pump pulse.}
    \label{fig:TR}
\end{figure}

To further illustrate the possibilities provided by the extracted single pulses, we investigate the temporal behavior of the switched and demagnetized areas of the sample using a single-shot time-resolved imaging technique first introduced by Vahaplar \textit{et al.}~\cite{Vahaplar} Such imaging of GdFeCo allows us to unambiguously identify ``time-zero'', i.e., the temporal overlap of the IR pump and visible probe pulses at the sample. The process of magnetization reversal takes place on the picosecond timescale and thus is too fast for us to measure given the pulse duration of the current probe laser~\cite{gd2}. However, the remagnetization dynamics of the material persist across a millisecond timescale, allowing us to find the temporal overlap between the (sub-)picosecond-long pump and nanosecond-long probe pulses through electronically stepping the arrival time of the probe pulse. Figure~\ref{fig:TR} shows the various stages in the process of IR-induced demagnetization. The images in the top row show the magnetization at the indicated time delay while the images in the bottom row show the domain pattern after the sample has reached equilibrium (after a few seconds). This comparison is necessary due to the randomized nature of the thermally-induced demagnetization pattern, which causes the pattern to change after the arrival of every pump. The magnetization reversal at the outer edge of the laser spot has already completed by the time the first image was taken at about \SI{5}{\nano \second} after the arrival of the pump pulse. In contrast, as soon as the pump pulse arrives, the center of the spot becomes fully demagnetized as indicated by the homogeneous grey colour. As the time-delay of the probe pulse increases, domains appear and grow in size. 

%%%%%%%%%%%%%%%%%%%%% fig 8 %%%%%%%%%%%%%%%%%%%%%%%%

It is thus interesting to see that while the real switching process in GdFeCo occurs \st{is} on the picosecond timescale~\cite{gd2}, the full recovery of the heavily demagnetized area in the center of the spot takes many orders of magnitude longer. Indeed, even after $\sim$~\SI{50}{\micro \second}, the domain pattern is still not the same as the relaxed state observed after the pulse, as shown in the bottom row. The final domain pattern is reached only milliseconds after the arrival of the pump. 

\subsection{Resonant phonon-induced magnetic switching in a doped yttrium-iron-garnet film}

The switching of magnetization in GdFeCo is thermally-driven and therefore independent of the pumping photon energy. In contrast, the IR range and intensity of the obtained pulses is much more relevant in the area of nonlinear phononics~\cite{nonlinear}. Strong resonant excitation of the phonon system is crucial for driving the anharmonic interactions that bring the system into a different equilibrium. Because potentially, and most interestingly, such excitation can lead to a permanent change of the order parameter, such as that recently demonstrated in a magnetic dielectric~\cite{stupa1}, standard stroboscopic techniques become unusable. Instead, single-shot techniques need to be used with sufficiently strong pump pulses.

\begin{figure}[t]
    \centering
    \includegraphics[width = 8cm]{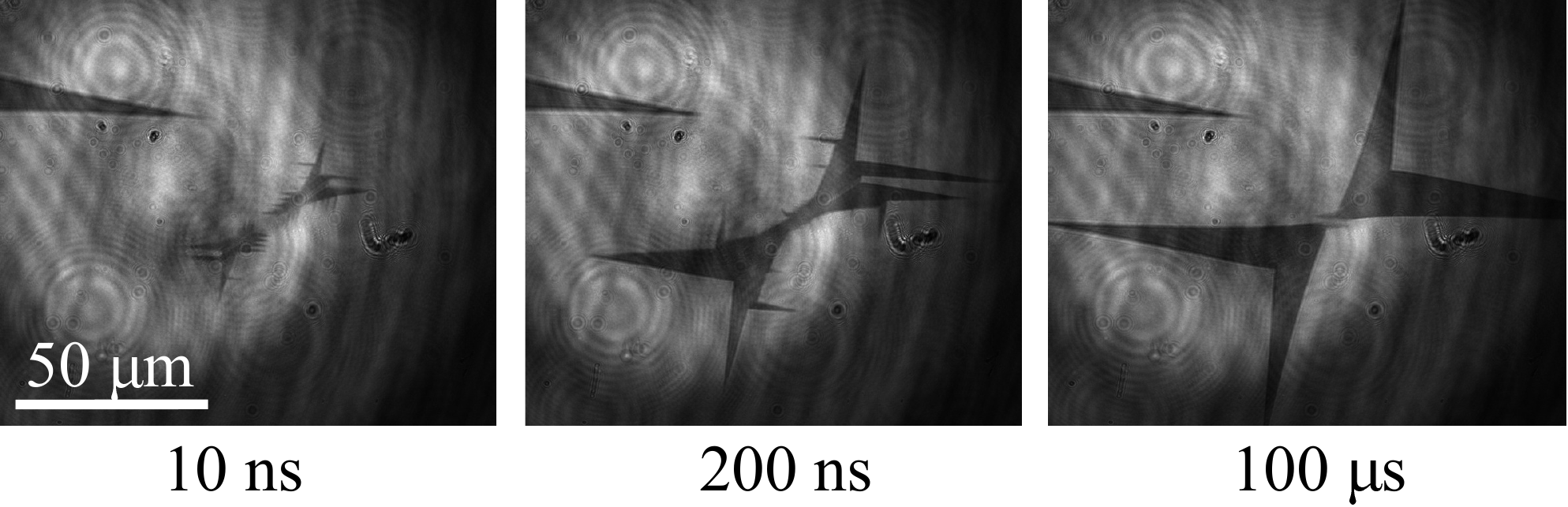}
    \caption{The effect of a cavity-dumped single micropulse on a transparent magnetic dielectric at three different time-delays. Ripples appear that seem to be moving outward from the center of the spot creating a bow-tie-shaped domain.}
    \label{fig:YIG}
\end{figure}

Here we apply our newly developed setup to benchmark the phonon-induced effects in transparent samples of magnetic dielectric doped yttrium-iron-garnet films. After resonantly pumping optical phonons at the wavelength of $\sim$~\SI{14}{\micro \meter}, a bow-tie shaped magnetic domain pattern appears that is stable for several milliseconds (Fig.~\ref{fig:YIG}). To compare, a similar pattern could only be observed at one of the FELIX beamlines after using an entire macropulse without any attenuation, to accumulate the effect of hundreds of micropulses. A single micropulse sliced from FELIX simply does not have enough energy to trigger the effect. The cavity-dumped micropulse from the FELICE beamline is, in contrast, capable of creating the magnetic domains, even with \SI{3}{\decibel} of attenuation. Moreover, the single-shot imaging with nanosecond time resolution as described in the previous section, we were able to resolve various phases of domain formation across different timescales. All details of this study will be published elsewhere~\cite{Maxime}.

%%%%%%%%%%%%%%%%%%%% fig 9 %%%%%%%%%%%%%%%%%%%%%%%%%

\section{Conclusions and Outlook}

In this work we have demonstrated the stable operation of a cavity-dumped free-electron laser in the mid- to far-infrared spectral range. We obtained pulses at \SI{10}{\hertz} repetition rate with energy exceeding \SI{130}{\micro \joule}, tunable in a wide spectral range. The duration of the Fourier-transform limited pulses could be adjusted between hundreds of femtoseconds to several picoseconds without disturbing the operation of the laser. Using the IR pulses as a pump, we have realized a two-color pump-probe magneto-optical imaging setup capable of tracking single-shot magnetization dynamics in a variety of materials. The capabilities of the obtained IR pulses were demonstrated by showing all-optical switching of magnetization in metallic ferrimagnet GdFeCo and magnetic dielectric YIG films. Using single pulses from the regular beam lines for comparison, we benchmarked the efficiency of switching, and time-resolved magneto-optical imaging was used to evaluate the dynamics of the magnetization generated by the single micropulse. 

To show the full capabilities of the setup it would be best to compare the performance of FELICE with and without the silicon slab across the whole operational range of FELICE. However, the FEL is optimized to operate in relatively small wavelength ranges compared to the full range. To explore the full operational range of the FEL, it has to be restarted, set up, and optimized many times, which would simply cost too much beamtime. Also the outcoupling Brewster window that is currently in place limits the range in which pulses can be dumped because it is made of ZnSe. To evaluate cavity dumping beyond the range of \SI{5}{\micro \meter} to \SI{20}{\micro \meter}, the window must be exchanged with another material that is transparent for such wavelengths.

While already demonstrating a very large gain in producing intense IR pulses, there are several aspects of the setup that can be improved. For example, a thinner silicon slab will have similar reflective properties but should have less negative impact on the cavity-buildup, resulting in higher intracavity power. Alternatively, it has recently been shown in Ref.~[\onlinecite{GeAu}] that gold-doped germanium represents another excellent material for cavity-dumping. Moreover, the addition of active cooling of the plasma switch should increase the reflectivity while simultaneously allowing it to handle larger intracavity powers. It is clear that with further optimization of the setup, one can further increase the energy of the cavity-dumped pulses. Clearly a potential is present to make the cavity dumped pulses two orders of magnitude more intense than single pulses delivered by the regular FEL beamlines.

\begin{acknowledgments}
The authors thank all technical staff of FELIX facility for technical support, and  A. Tsukamoto and T. Johansen for providing GdFeCo and YIG samples, respectively. This research has received funding from de Nederlandse Organisatie voor
Wetenschappelijk Onderzoek (NWO) (Contract nr. 680-91-131). 
\end{acknowledgments}

\section*{Data availability}

The data that support the findings of this study are available
from the corresponding author upon reasonable request.

\nocite{*}
\bibliography{FLC_Corr}

\end{document}